\documentclass[journal=nalefd,manuscript=letter]{achemso}
\usepackage[version=3]{mhchem}
\usepackage{amsmath}
\usepackage{amsfonts}
\usepackage{amssymb}
\usepackage{graphicx}

\author{Damon J. Carrad}
\author{Adam M. Burke}
\author{Roman W. Lyttleton}
\affiliation[University of New South Wales]{School of Physics,
University of New South Wales, Sydney NSW 2052, Australia}
\author{Hannah J. Joyce}
\author{Hark Hoe Tan}
\author{Chennupati Jagadish}
\affiliation[Australian National University] {Department of
Electronic Materials Engineering, Research School of Physics and
Engineering, The Australian National University, Canberra ACT 0200,
Australia}
\author{Kristian Storm}
\author{Heiner Linke}
\author{Lars Samuelson}
\affiliation[Lund University]{Solid State Physics/Nanometer
Structure Consortium (nmC@LU), Lund University S-221 00 Lund,
Sweden}
\author{Adam P. Micolich}
\email{adam.micolich@nanoelectronics.physics.unsw.edu.au}
\affiliation[University of New South Wales]{School of Physics,
University of New South Wales, Sydney NSW 2052, Australia}

\date{\today}

\title[] {Electron-beam patterning of polymer electrolyte films to make multiple nanoscale gates for nanowire transistors}

\begin{document}

\begin{abstract}
We report an electron-beam based method for the nanoscale patterning
of the poly(ethylene oxide)/\ce{LiClO4} polymer electrolyte. We use
the patterned polymer electrolyte as a high capacitance gate
dielectric in single nanowire transistors and obtain subthreshold
swings comparable to conventional metal/oxide wrap-gated nanowire
transistors. Patterning eliminates gate/contact overlap which
reduces parasitic effects and enables multiple, independently
controllable gates. The method's simplicity broadens the scope for
using polymer electrolyte gating in studies of nanowires and other
nanoscale devices.

\end{abstract}

\maketitle

Polymer electrolytes~\cite{KimAdvMat13} and III-V nanowire
transistors~\cite{SamuelsonMatsTod03, ThelanderMatsTod06} are two
exciting outcomes of recent research on nanoscale devices and novel
electronic materials. A polymer electrolyte typically consists of a
salt dissolved in a solid polymeric matrix, e.g., \ce{LiClO4} in
poly(ethylene oxide) (PEO)~\cite{ArmandAnnuRevMatSci86}; they are
commonly used as a gate dielectric in organic field-effect
transistors. The electric field resulting from a voltage applied to
the gate drives motion of \ce{Li+} and \ce{ClO4-} ions through the
polymer matrix to form an electric double layer (EDL) at the
gate/insulator and insulator/channel interfaces. EDL formation
effectively transfers the gate charge to $\sim1$~nm away from the
channel,~\cite{TakeyaAPL06} producing the high dielectric constants
and specific capacitances for which polymer electrolyte gate
dielectrics are known.~\cite{KimAdvMat13} The benefits of reduced
operating voltages~\cite{PanzerAPL05} and enhanced carrier
density~\cite{PanzerAFM06} that polymer electrolytes bring to
organic transistors have seen them applied to one-dimensional
nanomaterials also; first with carbon
nanotubes,~\cite{RosenblattNL02, SiddonsNL04, OzelNL05} and more
recently, with self-assembled InAs nanowires.~\cite{LiangNL12} The
latter is part of a broader quest to improve electrostatic gate
control in nanowire-based devices, both for fundamental transport
studies and potential nanowire device applications.

The first nanowire transistors were gated using a \ce{SiO2}-coated,
degenerately-doped Si substrate; though effective, this approach
provides no local control over carrier density.~\cite{DuanNat01}
Subsequent work led to patterned local gating of laterally-oriented
nanowires via electrodes both under~\cite{FasthNL05} and
over~\cite{PfundAPL06} the nanowire, and more recently, with a
concentric `wrap-gate'~\cite{StormNL12, DharaAPL11}. Wrap-gates
provide more homogeneous carrier depletion and better gate/channel
coupling~\cite{KhanalNL07}, give improved subthreshold
characteristics and reduced operating voltage,~\cite{ThelanderTED08,
StormNL12, DharaAPL11} and enable more controllable devices for
fundamental studies of 1D transport.~\cite{FordNL12, TianNL12,
vanWeperenNL13}

Liang and Gao's use of a PEO/\ce{LiClO4} polymer electrolyte gate
spin-coated over an InAs nanowire provides a simpler route to
lateral wrap-gated nanowire transistors;~\cite{LiangNL12} however, a
key limitation resides in a lack of methods for nanoscale patterning
of polymer electrolytes. Patterning the polymer electrolyte is
desirable to avoid it overlapping the source/drain contacts, which
can lead to parasitic capacitance, leakage currents and contact
corrosion~\cite{KimAdvMat13}. It also enables independent contacting
of multiple devices on the same chip. The micron-scale resolution of
established polymer electrolyte deposition methods, e.g., ink-jet
printing,~\cite{deGansAdvMat04, ChoNatMat08} injection into
microfluidic channels~\cite{OzelNL05} and
photolithography,~\cite{ChoiJMatChem10} presently limits the use of
polymer electrolytes in nanowire transistors, where $200$~nm -
$3~\mu$m channel lengths are typical. Here we report the development
of a process for electron-beam patterning of the PEO/\ce{LiClO4}
polymer electrolyte, and demonstrate the versatility it provides by
making nanoscale-patterned single and double electrolyte-gated
nanowire transistors. This nanoscale patterning capability enables
us to produce multiple independent devices, each with multiple
independently controllable electrolyte gates, on a single chip.

Electron Beam Lithography (EBL) is a widely used tool for nanoscale
patterning; it relies on using electron-induced chain
scission/crosslinking to locally alter the solubility of a polymeric
`resist' layer in a `developer' solution. PEO can be crosslinked by
exposure to energetic electrons, which makes these regions
comparatively insoluble in developers such as tetrahydrofuran,
methanol and \ce{H2O}; as such PEO is a negative-tone EBL resist,
though not widely used practically. Krsko {\it et al.} first
demonstrated EBL of PEO,~\cite{KrskoLangmuir03} with feature sizes
down to $\sim 200$~nm achieved soon
thereafter~\cite{HongLangmuir04}. These works used PEO with
molecular weights (MW) of $6.8$ and $200$~k without any added salts,
and $10$~keV electrons at doses $1-200$~C/m$^{2}$. In implementing
EBL-patterning of a polymer electrolyte there are some new concerns
that arise, e.g., whether the added salt either captures incident
electrons or adversely affects electron-induced crosslinking, and
whether the cross-linked PEO remaining after development has
sufficient ionic mobility to produce a functional device. While EBL
patterning of salt-doped PEO for nanoscale functional polymer
electrolyte gates has not been previously demonstrated, prior
research suggests its viability, e.g., electron-beam crosslinking
has been used to enhance ionic conductivity in solid polymer
electrolytes for battery applications.~\cite{UchiyamaSSI09,
UenoJPS11}

\begin{figure}[t]
\includegraphics[width=8cm]{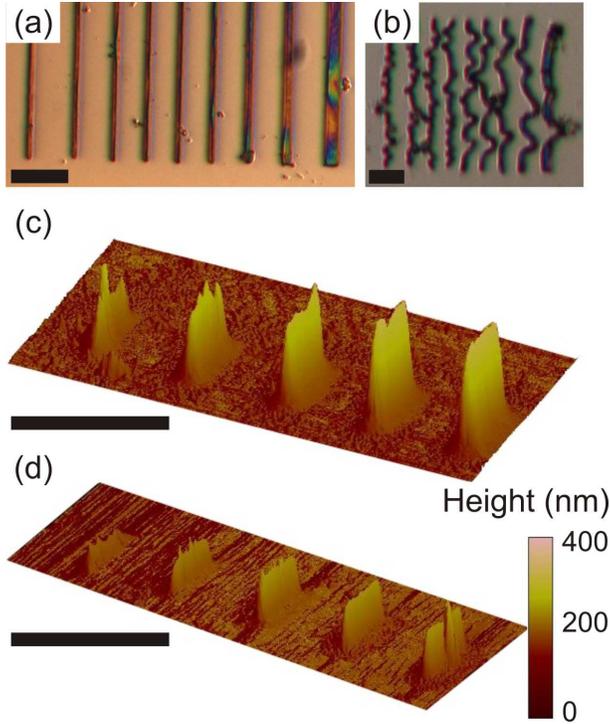}
\caption{(a/b) Optical micrographs of PEO/\ce{LiClO4} patterned by
EBL into 50~$\mu$m-long lines of defined width $w_{d} =$ $100$,
$150$, $200$, $300$, $400$, $500$, $750$, $1000$ and $2000$~nm. The
dose in (a) was $4$~C/m$^2$, and the measured line width saturates
below defined widths of $500$~nm. The dose in (b) was $1$~C/m$^2$:
low doses occasionally resulted in pattern distortion during
development. (c/d) Atomic force micrograph of $4~\mu$m-long lines of
defined width $100$~nm, with doses of $0.5$, $1$, $2$, $3$ and
$4$~C/m$^2$ with polymer:salt ratios of (c) $10:1$ and (d) $2.4:1$.
All patterns are on Si substrates. The black scale bars represent
$15~\mu$m. Cross-sectional line scans of (c/d) are shown in Fig.~S1
of the supplementary information.}
\end{figure}

Polymer electrolytes were formed by mixing PEO (Aldrich, MW $100$~k)
and \ce{LiClO4.3H2O} (Aldrich) in polymer:salt ratios of $10:1$,
$8:1$ and $2.4:1$ by sonication in $10$~mL of methanol. The
resulting mixture was left standing at room temperature overnight to
precipitate out large particulates, with the supernatant used for
deposition. The solution was spun onto the sample at $4000$~rpm for
$60$~s and the sample was then baked on a hot-plate at $90^\circ$C
for $30$~mins. The resulting film was EBL patterned using either an
FEI Sirion for preliminary experiments (Fig.~1), or a Raith
$150$-Two for nanowire device fabrication (Figs~2-4). Patterning was
performed using a $5$~kV accelerating voltage and beam currents of
$20-25$~pA under high vacuum. The patterned films were developed in
deionized water at room temperature for $\sim30$~s and dried with
\ce{N2} gas.

The optical micrograph in Fig.~1(a) shows $10:1$ PEO/\ce{LiClO4}
patterned at an electron dose $d = 4$~C/m$^2$ into lines with
different defined line widths, aimed at establishing the
patternability of PEO/\ce{LiClO4} films using EBL, and an initial
assessment of resolution limit. Pattern broadening is common for
negative tone EBL resists and is caused by the proximity effect --
the same physics produces the undercut profile for positive tone EBL
resists such as polymethylmethacrylate (PMMA).~\cite{MurataJAP81,
ShimizuRPP92} This means that the final pattern dimensions can be
significantly greater than the region scanned by the electron beam.
Thus, two line widths are important: the `defined' line width,
$w_{d}$, as written by the electron beam, and the measured line
width after development, $w_{m}$, which we take as the full width at
half maximum determined by atomic force microscopy. The proximity
effect can also result in $w_{m}$ depending strongly on dose: This
is suggested in Fig.~1(a), where there is a clear difference in
measured line width $w_{m}$ for wider lines ($w_{d} = 750 -
2000$~nm) but $w_{m}$ saturates for $w_{d} \leqslant 500$~nm with $d
= 4$~C/m$^2$. We used atomic force microscopy to study the effects
that electron dose, polymer:salt ratio and substrate material have
on the shape and dimensions of structures remaining after
development.

Figures~1(c/d) show $w_{d} = 100$~nm lines exposed at $d =
0.5-4$~C/m$^{2}$ for two polymer:salt ratios $10:1$ (Fig.~1(c)) and
$2.4:1$ (Fig.~1(d)). Focussing first on dose, in Fig.~1(c) $w_{m}$
decreases continuously from $1.2~\mu$m to $820$~nm as $d$ is reduced
from $4$~C/m$^2$ to $0.5$~C/m$^2$. This is expected for proximity
effect controlled line-width. Looking more closely at the
base-broadening, the width at the substrate can be up to $2 \times
w_{m}$, but the profiles in Fig.~1(c) show that most of the
broadening occurs for heights $< 50$~nm above the substrate surface.
This suggests the broadening arises due to surface effects, and as
such, the base width may be controllable with surface treatments; we
will address this in future work. The reduction of $w_{m}$ with
lower $d$ would imply that minimizing $d$ is most optimum but there
are two additional factors that weigh against this: line height and
surface adhesion. First and foremost, the line height $h$ in
Fig.~1(c) decreases with $d$, from $h = 350$~nm at $d = 4$~C/m$^2$
to $150$~nm at $0.5$~C/m$^2$. This aspect is particularly crucial to
the application in nanowire transistors because for a nanowire of
radius $R$ the electrolyte gate needs to have $h > 2R$ after
development to cover the nanowire without discontinuity. This
height-dose relationship sets an absolute minimum dose for
patterning. Additionally, surface adhesion of the patterned PEO
encourages further increases in dose. Figure~1(b) shows an issue
that frequently arises for $d~\leqslant~1$~C/m$^2$. Here the exposed
PEO is insufficiently crosslinked to prevent detachment of defined
lines from the substrate; the significant line deformation arises
from unrestrained swelling of the PEO due to \ce{H2O} uptake during
development.~\cite{KrskoLangmuir03} This problem becomes
particularly prevalent for $w_{d} < 100$~nm.

Small changes in polymer:salt ratio, e.g., from $10:1$ to $8:1$,
produced little appreciable pattern change, but as Fig.~1(d) shows,
$w_{m}$ and $h$ reduce substantially for a larger increase in salt
content to $2.4:1$. This suggests that ionic capture of incident
electrons at the expense of crosslinking occurs; this can be
mitigated to some extent by an increase in dose. Comparing the
left-most line in Fig.~1(c) with the right-most line in Fig.~1(d)
suggests that a $5\times$ increase in salt content requires a
$8\times$ increase in dose. Finally, we find that $w_{m}$ for a
given $w_{d}$ and $d$ combination improves by up to $200$~nm on
moving from a Si substrate with native oxide only to an n$^{+}$-Si
substrate capped with $100$~nm thermal oxide and $10$~nm \ce{HfO2}
deposited by atomic layer deposition. The line detachment effect in
Fig.~1(b) also becomes less prevalent for the
\ce{HfO2}/\ce{SiO2}-capped substrates used in nanowire processing.
These improvements may be due to improved adhesion of PEO to the
substrate surface and modification of the electron beam interaction
volume due to the layered oxide structure,~\cite{AizakiJVST79}
combined with the much higher electrical conductivity of the
underlying Si. There may be scope for further improvement in
resolution, e.g., with added reagents for controlling crosslinking.
A line width of $500~-~1000$~nm is sufficient to gate the
$3~-~6~\mu$m long InAs nanowires we use here without
electrolyte/contact overlap, so we leave this further process
development for future work and now turn to the nanowire devices.

\begin{figure}[t]
\includegraphics[width=17cm]{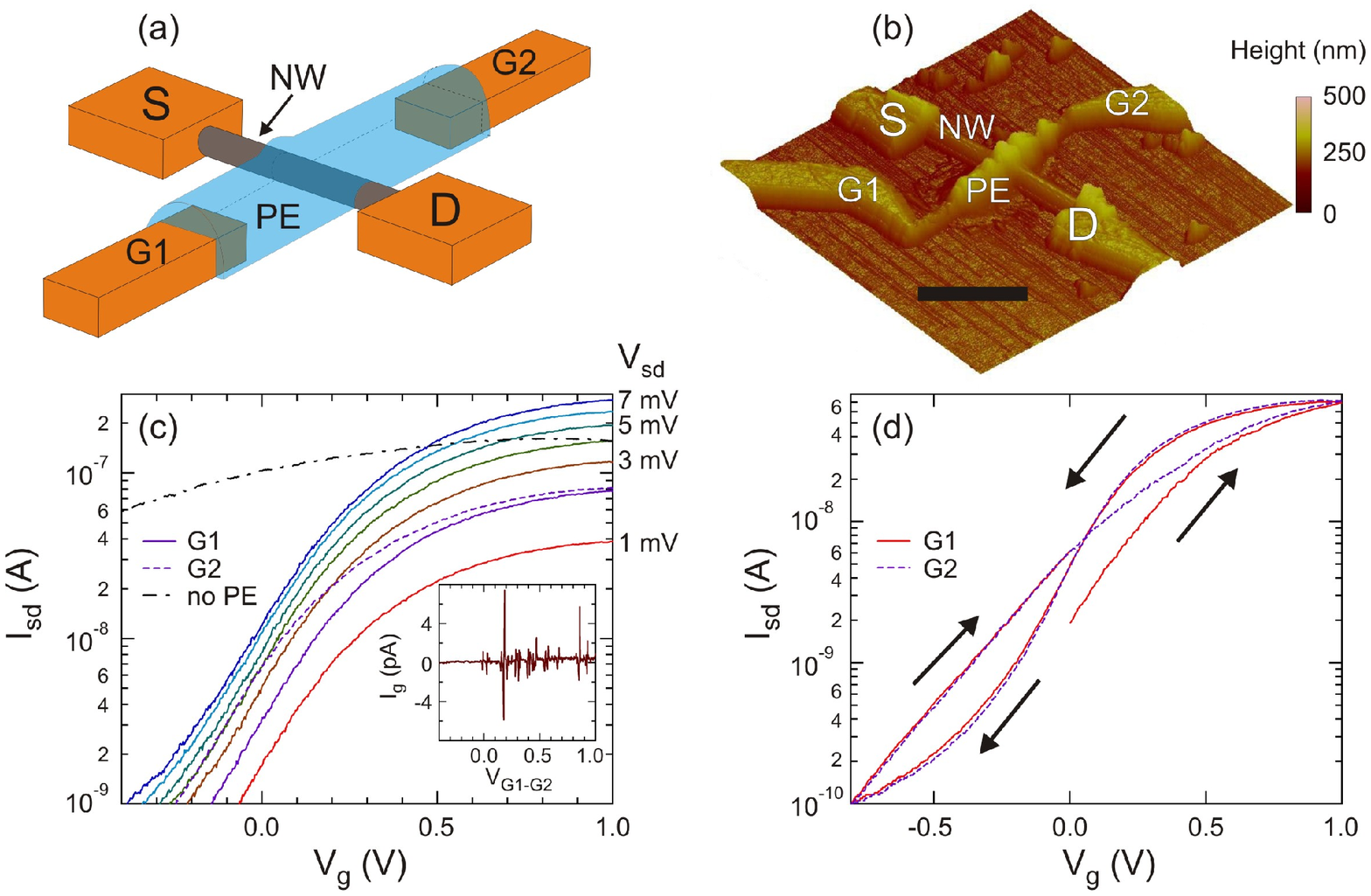}
\caption{(a) Schematic and (b) atomic force micrograph of a polymer
electrolyte-gated nanowire transistor. The components are labeled:
source (S), drain (D), nanowire (NW), polymer electrolyte (PE) and
gate electrodes (G1 and G2). The black scale bar represents a
horizontal distance of 3 $\mu$m. (c) Source-drain current $I_{sd}$
vs gate voltage $V_{g}$ with $V_{g}$ applied to G1 (solid lines) and
G2 (dashed line) of a device with $S_{G2} = 2~\mu$m, and G1 for a
device with no PEO/\ce{LiClO4} (dot-dash line). Traces are shown for
$V_{sd} = 1 - 7$~mV. Data was obtained with a $V_g$ sweep rate of
$5$~mV/s from positive to negative. Inset to (c) shows the current
$I_{g}$ flowing between G1 and G2 when a voltage $V_{G1-G2}$ is
applied between them. (d) Gate hysteresis at a sweep rate of sweep
rate of $5$~mV/s for $V_{g}$ applied to G1 (solid line) and G2
(dashed line) for a device with $S_{G2} = 4~\mu$m.}
\end{figure}

Figures~2(a/b) show a nanowire transistor incorporating a single
polymer electrolyte gate (PE) connected to two Ni/Au gate electrodes
(G1 and G2). The second electrode was used to test whether the
electrode-nanowire separation, $S_{G1}$ or $S_{G2}$, influences PE
gate operation; electrodes G1 and G2 are separated from the nanowire
by $S_{G1} = 1~\mu$m (fixed) and $S_{G2} = 1-4~\mu$m (varied between
devices), respectively. The PE gate has a polymer:salt ratio of
$10:1$ and was written with $w_{d} = 100$~nm and $d = 1$~C/m$^{2}$,
giving a strip with $w_m = 650$~nm and $h \sim 130$~nm. Figure~2(c)
shows the source-drain current $I_{sd}$ versus PE gate voltage
$V_{g}$ for seven different source-drain biases $V_{sd}$ between $1$
and $7$~mV with $V_{g}$ applied to G1 (solid lines) and at $V_{sd} =
2$~mV with $V_{g}$ applied to G2 (dashed line). In all experiments,
the electrode that did not have $V_{g}$ applied was kept at ground -
however there was no major difference to the transfer
characteristics if this electrode was floated (see Supplementary
Figure~S2(a)). There is hysteresis in the gate characteristics, as
we show in Fig.~2(d) and discuss further below. Hence for the data
in Fig~2(c) we only show data obtained for one sweep direction: from
positive $V_g$ towards more negative $V_g$. Considering data for
$V_{g}$ applied to G1 first, we obtain a subthreshold swing of
$271$~mV/decade for data in Fig.~2(c). Across $12$ devices studied
so far with $10:1$ polymer:salt ratio (a total of $22$ working
gates) we obtain an average subthreshold swing
$307~\pm~33$~mV/decade. The average threshold voltage was $+0.16 \pm
0.06$~V at $V_{sd} = 2$~mV across the $12$ devices. We now look at
the influence of the separation between the gate electrode and the
nanowire on the transistor characteristics. For the device measured
in Fig~.2(d), $S_{G1} = 1~\mu$m and $S_{G2} = 2~\mu$m. Despite this
difference, the gate characteristics in Fig.~2(c) are very similar
with almost identical subthreshold swing. We find this same
behaviour across many devices where $S_{G2}$ ranges from $1~\mu$m to
$4~\mu$m (see Supplementary Figure S2(b)). The lack of dependence of
the sub-threshold swing on gate electrode to nanowire separation is
not surprising; for an ideal EDL, $V_g$ drops across the
nanowire/electrolyte and electrode/electrolyte interfaces, not
across the electrolyte itself. The result is that the steady state
gate capacitance -- and thereby the subthreshold swing -- is
independent of the electrode-nanowire separation. Note that the
polymer electrolyte is not electronically conductive; Fig.~2(c)
(inset) shows a plot of current through the PE gate $I_{g}$ versus
potential difference between electrodes G1 and G2 $V_{G1-G2}$
demonstrating a negligible electronic conductivity despite a
significant ionic conductivity.

Gate hysteresis is a common issue for transistors incorporating
polymer electrolyte gate dielectrics. It normally arises due to the
finite ionic mobility of the polymer electrolyte, since ions need to
drift through the polymer to re-establish electrostatic equilibrium
at the EDLs when the voltage on the gate electrode is altered. This
hysteresis will depend on properties of the polymer electrolyte, but
also on the distance between the gate electrode and transistor
channel. On its own, the delay imposed by ion migration means that
$I_{sd}$ for sweeps from positive (negative) to negative (positive)
gate voltages will be higher (lower) than otherwise expected,
producing a counter-clockwise hysteresis loop. Figure~2(d) shows
extended $I_{sd}$ vs $V_g$ traces for G1 (solid red line) and G2
(dashed purple line) for a device with $S_{G1} = 1~\mu$m and $S_{G2}
= 4~\mu$m. Two interesting features are evident. First, neither gate
trace follows a simple, counter-clockwise cyclical loop; instead
they take a `figure of 8' form that indicates possible additional
contributions to the hysteresis. One additional contribution may be
charge trapping by nanowire surface states\cite{LindNL06,
ThelanderTED08}, which would depend on the exact nature of the
InAs/PEO interface. Second, aside from the `virgin' behavior in the
initial positive ramp of G1, the hysteresis traces for G1 and G2 are
identical, despite the factor of 4 difference between $S_{G1}$ and
$S_{G2}$. While the gate response is not identical for all devices,
there is no clear relation between the magnitude of the hysteresis
and $S_{G2}$. This also points to contributions other than ionic
mobility to the hysteresis, and suggests that these other
contributions are dominant. Indeed, much smaller hysteresis loops
are typically seen in organic transistor and carbon nanotube devices
with PEO/\ce{LiClO4} gate dielectrics\cite{OzelNL05, PanzerAPL06}.
We characterise the hysteresis further in the Supplementary
Information, but most notably, the hysteresis can be reduced
significantly by sweeping over a smaller gate range and/or at a
lower rate. Determining the relative contributions of ionic
mobility, surface states and other possible contributions to the
gate hysteresis is beyond the scope of this work, but would be an
interesting subject for future studies.

We now consider the effect of increased salt content on the device
structure in Fig.~2(a/b). The first place where this presents an
effect is in device fabrication. Unlike the test structures in
Fig.~1, here we need to precisely align the PE gate to the gate
electrodes and nanowire, and this is done by briefly viewing metal
alignment markers on the substrate immediately prior to EBL
patterning to ensure correct pattern alignment. Increasing the
polymer:salt ratio to $8:1$ makes the PEO/\ce{LiClO4} film opaque to
the electron beam, resulting in difficulties in pattern alignment
and thereby dramatically reducing device yield. Interestingly,
increasing the polymer:salt ratio to $2.4:1$ returns some of the
PEO/\ce{LiClO4} film's transparency to an electron beam -- we
explain this below. The second place where we see a salt
concentration effect is in the electrical characteristics. Despite
reduced yield, we successfully measured four devices at $8:1$
obtaining an average subthreshold swing of $286~\pm~45$~mV/decade
from eight gates, and four devices at $2.4:1$ obtaining an average
subthreshold swing of $431~\pm~53$~mV/decade from eight gates
measured. Both the maximal electron beam opacity of the
PEO/\ce{LiCLO4} film and maximal subthreshold swing at intermediate
polymer:salt ratio can be explained by the `ionic conductivity peak'
observed as a function of salt
concentration;~\cite{BruceJChemSocFarTran93} this peak typically
occurs at $8:1$.~\cite{Fullerton-ShireyMacroMol09} Thus we have
concluded that a $10:1$ polymer:salt ratio offers the best
compromise between patternability and device performance for the
remainder of this work.

Comparing the performance of our PE gated devices to other nanowire
transistors, our typical subthreshold swing of $\sim 300$~mV/decade
compares very favorably with substrate-gated nanowire transistors,
where subthreshold swings of order $1-4$~V/decade are
typical.~\cite{StormNL12} The performance is also competitive with
metal/oxide wrap-gated nanowire transistors, where subthreshold
swings typically range from 100 to 750~mV/decade.~\cite{StormNL12,
RehnstedtTED08, FrobergEDL08, RehnstedtEL08, ThelanderTED08,
TanakaAPEX10} This is particularly impressive as the polymer
electrolyte does not completely wrap around the nanowire in our
devices, unlike in Ref.~\cite{LiangNL12} where a HF etch was used to
`undercut' the nanowire to provide access for the PEO/\ce{LiClO4}
film. This undercut etch was impractical to implement here as our
\ce{HfO2} cap layer is much more resistant to HF etching than
\ce{SiO2};~\cite{StormNL12} this step could easily be implemented
for substrates with a thermally grown \ce{SiO2} layer alone.

\begin{figure}[t]
\includegraphics[width=8cm]{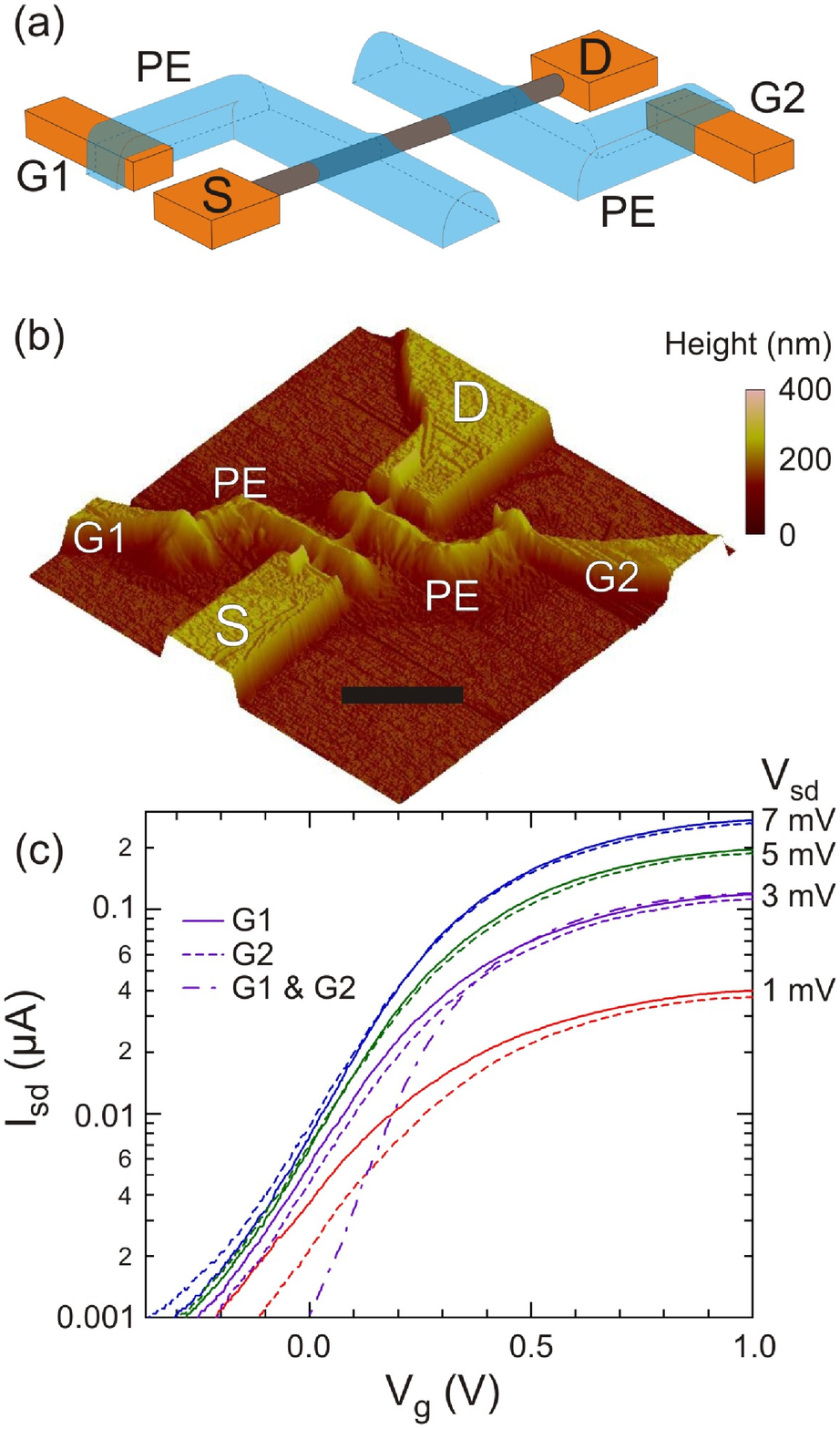}
\caption{(a) Schematic and (b) atomic force micrograph of a dual
PE-gated nanowire transistor. The source (S), drain (D), polymer
electrolytes (PE) and gate electrodes (G1 and G2) are labeled. The
black scale bar represents a horizontal distance of 3 $\mu$m. (b)
Source-drain current $I_{sd}$ vs gate voltage $V_{g}$ for the device
in (a) for G1 (solid lines), G2 (dashed lines), and G1 and G2 biased
together (dot-dash line). Traces are shown for $V_{sd} = 1, 3, 5,
7$~mV and obtained with a $V_g$ sweep rate of $5$~mV/s.}
\end{figure}

A concern that could be raised is that closely positioned, biased
metal electrodes can also influence nanowire
conduction.~\cite{RoddaroNL11} Since we observed evidence for some
electron capture by \ce{Li+} ions at the patterning stage, the
question that naturally follows is: To what extent does the direct
coupling of the metal electrodes to the nanowire contribute to
modulation of $I_{sd}$ with $V_{g}$? For example, is it that
\ce{Li+} ions are neutralized by incident electrons during EBL such
that ionic conduction is a co-contributor with electrostatic
repulsion via the gate electrodes, rather than the dominant
contributor to channel depletion? The black dot-dash trace in
Fig.~2(c) shows the characteristics for a device like that in
Fig.~2(a/b), but without any PEO/\ce{LiClO4}. At $V_{g} = -0.4$~V
the bare electrode has only reduced $I_{sd}$ by a factor of $2$
compared to factor of $>10^{2}$ for the PE-gated device. Pinch-off
can be achieved with a bare electrode, but it requires $V_{g} \sim
-3$~V with $S_{G2} = 1~\mu$m and a much more negative $V_{g}$ at
greater electrode-nanowire separations. This is expected, since
there is no EDL formation for bare electrode gating. The behavior of
the bare electrode suggests that EDL formation is the dominant
contributor to channel depletion in devices with a polymer
electrolyte, despite any ionic mobility loss or neutralization that
may arise from the EBL process. We confirm this via one final test
with our dual PE-gated devices, which we now discuss.

Figures~3(a/b) show a PE-gated device with two independent gates.
Here, we have located the electrodes G1 and G2 such that their
direct electrostatic coupling to the nanowire is screened by the
source/drain contacts. This ensures that all depletion in this
device arises from EDL formation by ion migration in the polymer
electrolyte, which had polymer:salt ratio $10:1$. Figure~3(c) shows
$I_{sd}$ versus $V_{g}$ for G1 with G2 grounded (solid lines) and G2
with G1 grounded (dashed lines) for several different $V_{sd}$. The
characteristics for the two gates are similar, with G1 and G2 giving
subthreshold swings of $332$ and $321$~mV/decade, respectively.
These values are within error for the average values obtained from
the single gate devices; this demonstrates that direct coupling to
the metal electrodes makes a negligible contribution to depletion
and that consistent performance can be obtained from our PE gate
structures. The dotted line in Fig.~3(c) shows the characteristics
obtained when $V_{g}$ is applied to both G1 and G2 simultaneously.
The subthreshold swing improves to $192$~mV/decade, suggesting that
performance gains might be achieved in our single PE-gate
transistors by careful adjustment of gate width.

\begin{figure}[t]
\includegraphics[width=17cm]{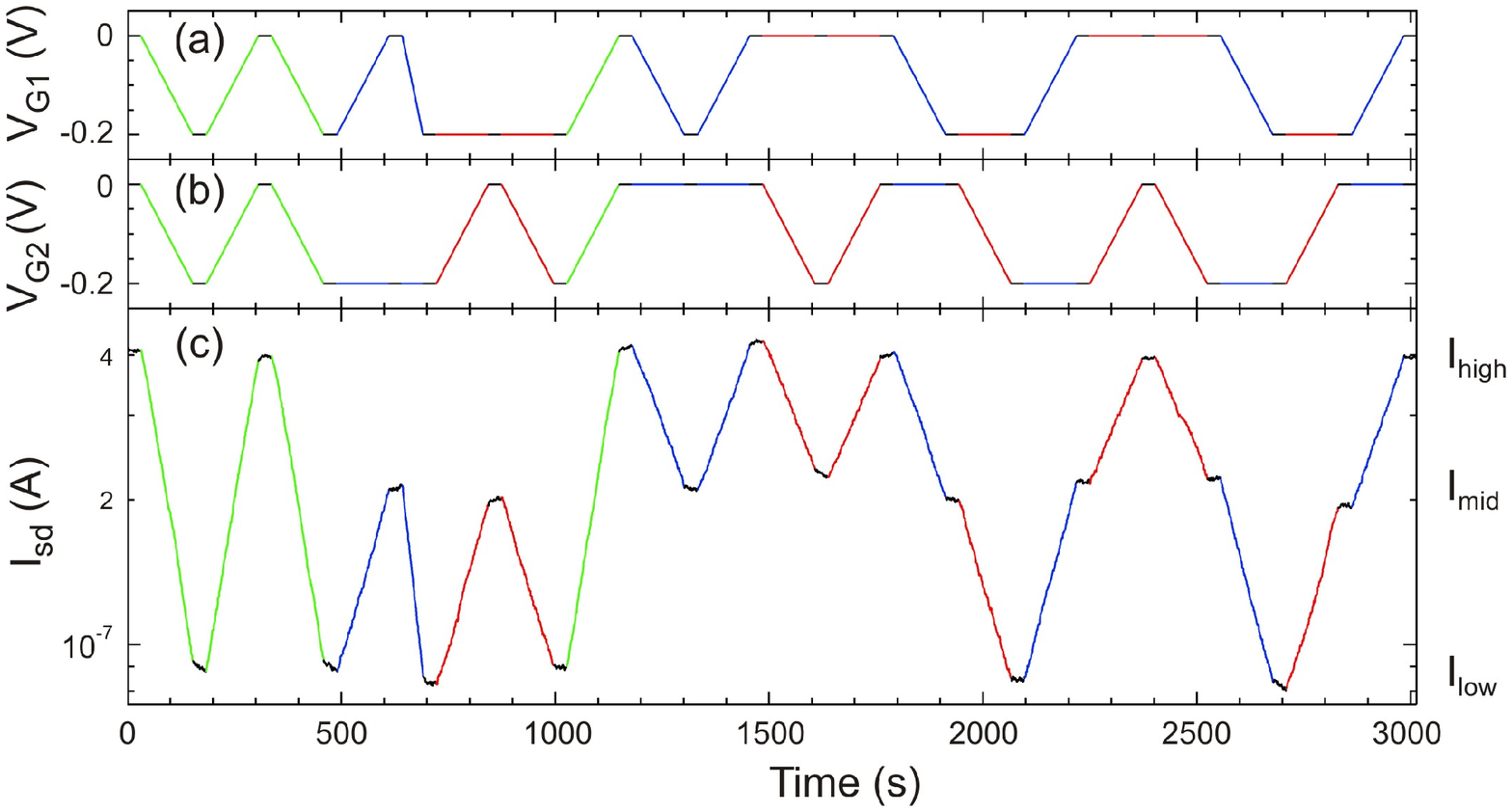}
\caption{(a) $V_{g}$ on G1, (b) $V_{g}$ on G2 and (c) $I_{sd}$ vs
time $t$ with constant at $V_{sd} = 42$~mV demonstrating independent
operation of the gates on the dual polymer electrolyte-gated
nanowire device. To highlight the actions, trace segments are
colored: green -- G1 and G2 swept together, blue -- G1 swept with G2
fixed, red -- G2 swept with G1 fixed. The stability of the resulting
$I_{sd}$ was monitored for 30 s at the end of each sweep in the
program (black sections). Gates were swept at $2$~mV/s.}
\end{figure}

To better assess the control, balance and temporal stability of
these gates, and gauge the potential for making more complex
devices, e.g., a single electron transistor, we performed a more
in-depth study of the two gates in our dual PE-gate transistor
device when used within an operating range with relatively low
hysteresis. This involved taking the device in Fig.~3(b) through a
`program' where G1 and G2 were swept together or separately between
two pre-defined voltages $V_A = -200$~mV and $V_B = 0$~V. The
program for G1 and G2 versus time $t$ is shown in Figs.~4(a) and
(b), respectively, with the $I_{sd}$ response plotted in Fig. 4(c).
The full program takes 50 min, and the program is paused after each
gate sweep to check stability for 30 s (black segments). The program
produces an $I_{sd}$ that alternates between three distinct current
states $I_{high}$, $I_{mid}$ and $I_{low}$, which correspond to
three gate configurations: $V_{G1}$ = $V_{G2}$ = $V_B$; $V_{G1}$ =
$V_A$, $V_{G2}$ = $V_B$ or $V_{G1}$ = $V_B$, $V_{G2}$ = $V_A$, and
$V_{G1}$ = $V_{G2}$ = $V_A$. The fact that $I_{sd} = I_{mid}$ for
both $V_{G1} = V_A$, $V_{G2} = V_B$ and $V_{G1} = V_B$, $V_{G2} =
V_A$ highlights the strong electrical balance between the two
nominally identical patterned PE gates. Further, the consistent
return to $I_{high}$, $I_{mid}$ and $I_{low}$ across the program in
Fig. 4 demonstrates the stability and low gate drift of this device
within this operating region.

One could equally view Fig.~4 as a demonstration of two-gate logic,
where we set input states [1, 1] ($V_{G1} = V_{G2} = V_B$), [1, 0]
($V_{G1} = V_B$, $V_{G2} = V_A$), [0, 1] ($V_{G1} = V_A$, $V_{G2} =
V_B$) and [0, 0] ($V_{G1} = V_{G2} = V_A$) giving either AND or OR
operations as output if the threshold is set above or below
$I_{mid}$, respectively. While logic is in principle possible in
this device, the time response of the polymer electrolyte gates is
insufficient to be competitive for applications. The data in Fig.~4
represents the fastest operation we can presently achieve, i.e., a
few mV/s, without compromising on stability and reproducibility in
$I_{sd}$ of the logic states. For faster sweeps, the current at each
state is less stable over the 30~s period and the value of the
current at each state varies throughout the program. This makes each
state less distinct and limits the practical switching speed to less
than 1~Hz. There is some scope for improving the switching speed in
future work by passivation of surface states, or engineering of the
polymer electrolyte, e.g. by optimizing PEO/\ce{LiClO4} salt content
or adding plasticisers/nanoparticles.~\cite{CroceNat98, QianJSSE01,
WangJNanosciNanotech05} While switching speeds from $1 - 100$~Hz
have been obtained in solid polymer electrolytes such as
PEO/\ce{LiClO4}, speeds up to $10$~kHz can be obtained by moving to
a special class of polymer electrolytes known as `ion
gels'.~\cite{KlingshirnChemMat04, SusanJACS05, KimAdvMat13,
ChoNatMat08, LeeJACS07} These consist of an ionic liquid, i.e., a
room-temperature molten salt, dispersed in a gel matrix typically
formed using a block copolymer. A common example is
1-ethyl-3-methylimidazolium bis(trifluoromethylsulfonyl)imide
([EMIM][TFSI]) in poly(styrene - block - ethylene oxide - block -
styrene).~\cite{ChoNatMat08} A first step may be to attempt EBL
patterning of PEO-containing ion gels;~\cite{LeeJACS07,
ChoiJMatChem10, KlingshirnChemMat04} however, it is possible these
would suffer the same resolution difficulties described above.
Another candidate may be PMMA-based ion gels:~\cite{ChoNatMat08,
SusanJACS05} PMMA is a high resolution negative-tone resist under
very high electron doses, with feature sizes as small as $150$~nm
reported~\cite{ZailerSST96}. This could potentially solve both the
resolution and switching speed problems encountered here, and lead
to nanoscale electrolyte gates with high resolution, ionic
conductivity and switching speed. Nevertheless, the patterned
PEO/\ce{LiClO4} polymer electrolyte is an effective gate dielectric
for applications where a strong field is required but fast switching
speeds are not.

In conclusion, we have demonstrated electron-beam patterning of
PEO/\ce{LiClO4} electrolyte that allows gating of individual InAs
nanowires with single or multiple independently controllable gates
without gate/contact overlap. The electrolyte facilitates strong
gate-channel coupling; the subthreshold swing of our devices is
comparable to that of wrap-gated devices and tends to improve with
greater channel coverage. The fabrication of these devices is
simpler than for wrap-gated devices: only one extra EBL step is
required compared to traditional substrate-gated nanowire
transistors. Our dual PE-gated devices exhibit independent gate
control and ability to perform basic logic operations.

{\bf Materials and methods.}

{\bf Fabrication} Nanowire devices were fabricated from 3~-~6 $\mu$m
long, 50~nm diameter InAs nanowires grown by MOCVD. Devices were
fabricated on $0.001 - 0.005~\Omega$.cm As-doped $(100)$ Si wafer
(Silicon Valley Microelectronics) with a $100$~nm thermal oxide and
an additional $10$~nm \ce{HfO2} layer deposited by atomic layer
deposition. This wafer was prepatterned with Ti/Au interconnects and
EBL alignment structures before being divided into smaller `chips'
on which nanowire transistors were made. Nanowires were deposited by
dry transfer using lab wipe. Source, drain and gate electrodes were
then defined by EBL using a Raith 150-two system. The EBL resist was
a $5\%$ solution of $950$k MW PMMA in anisole (Microchem) deposited
by spin coating at $5000$~rpm followed by a $5$~min hotplate bake at
$180^{\circ}$C, developed using a $1:3$ mixture of
methylisobutylketone in 2-propanol. The electrodes consisted of
25~nm Ni and 75~nm Au deposited by thermal evaporation, immediately
after a 120~s (NH$_4$)$_2$S$_x$ contact passivation step at
$40^{\circ}$C.\cite{SuyatinNanotech07} Lift-off was performed
overnight in N-methyl-2-pyrrolidone at $80^{\circ}$C. Following
lift-off, a PEO/\ce{LiClO4} film was spin-coated, baked and
patterned by EBL as described in the main text to produce completed
devices, which were then packaged in LCC20 ceramic chip carriers
(Spectrum) and bonded using an Au ball bonder (Kulicke \& Soffa
4500). Atomic Force Microscopy (AFM) studies were performed prior to
packaging using a Dimension DI-3000 AFM in tapping mode using Veeco
OTESPA7 probes. AFM was performed in cleanroom ambient atmosphere
(temperature $20^{\circ}$C and relative humidity $50 - 60\%$).

{\bf Electrical Characterization}

All electrical characterization presented here was performed at room
temperature and atmosphere. The source-drain current was measured
using a Stanford Research Systems SRS830 lock-in amplifier with an
a.c. excitation $V_{sd} = 1-50$~mV applied at a frequency of $73$~Hz
using the internal oscillator. Gate electrodes were biased to
$V_{g}$ using Yokogawa GS200 or Keithley 2400 voltage sources with
built-in current monitoring for tracking the gate leakage current.
The gate current $I_{g}$ in the inset of Fig.~2(c) was monitored by
a Keithley 6517A electrometer.

Devices were stored in the dark in vacuum between measurements to
preserve the quality of the ohmic contacts~\cite{SuyatinNanotech07}
and polymer electrolytes. Under these conditions, device
characteristics typically remained reproducible for a period of 3 -
4 months before beginning to degrade (see Supplementary Figure S4).

{\bf Supporting Information.} Additional supporting data on
line-width studies as well as device characteristics, hysteresis
and longevity. This material is available free of charge via
the Internet at http://pubs.acs.org.

{\bf Corresponding author.} *E-mail:
adam.micolich@nanoelectronics.physics.unsw.edu.au

\acknowledgement

This work was funded by the Australian Research Council (ARC),
Nanometre Structure Consortium @ Lund University (nmC@LU), Swedish
Research Council (VR), and Knut and Alice Wallenberg Foundation
(KAW) APM acknowledges an ARC Future Fellowship (FT0990285). We
thank D. Liang and X.P.A. Gao for helpful discussions, and D.
Alvares, P.-H. Prevot and F. Ladouceur for assistance with initial
polymer electrolyte development. This work was performed in part
using the NSW and ACT nodes of the Australian National Fabrication
Facility (ANFF).

\bibliography{Carrad_NL13bib}

\providecommand*\mcitethebibliography{\thebibliography}
\csname @ifundefined\endcsname{endmcitethebibliography}
  {\let\endmcitethebibliography\endthebibliography}{}
\begin{mcitethebibliography}{49}
\providecommand*\natexlab[1]{#1}
\providecommand*\mciteSetBstSublistMode[1]{}
\providecommand*\mciteSetBstMaxWidthForm[2]{}
\providecommand*\mciteBstWouldAddEndPuncttrue
  {\def\EndOfBibitem{\unskip.}}
\providecommand*\mciteBstWouldAddEndPunctfalse
  {\let\EndOfBibitem\relax}
\providecommand*\mciteSetBstMidEndSepPunct[3]{}
\providecommand*\mciteSetBstSublistLabelBeginEnd[3]{}
\providecommand*\EndOfBibitem{}
\mciteSetBstSublistMode{f}
\mciteSetBstMaxWidthForm{subitem}{(\alph{mcitesubitemcount})}
\mciteSetBstSublistLabelBeginEnd
  {\mcitemaxwidthsubitemform\space}
  {\relax}
  {\relax}

\bibitem[Kim et~al.(2013)Kim, Hong, Xie, Lee, Zhang, Lodge, and
  Frisbie]{KimAdvMat13}
Kim,~S.~H.; Hong,~K.; Xie,~W.; Lee,~K.~H.; Zhang,~S.; Lodge,~T.~P.;
  Frisbie,~C.~D. \emph{Advanced Materials} \textbf{2013}, \emph{25},
  1822--1846\relax
\mciteBstWouldAddEndPuncttrue
\mciteSetBstMidEndSepPunct{\mcitedefaultmidpunct}
{\mcitedefaultendpunct}{\mcitedefaultseppunct}\relax
\EndOfBibitem
\bibitem[Samuelson(2003)]{SamuelsonMatsTod03}
Samuelson,~L. \emph{Materials today} \textbf{2003}, \emph{6}, 22 -- 31\relax
\mciteBstWouldAddEndPuncttrue
\mciteSetBstMidEndSepPunct{\mcitedefaultmidpunct}
{\mcitedefaultendpunct}{\mcitedefaultseppunct}\relax
\EndOfBibitem
\bibitem[Thelander et~al.(2006)Thelander, Agarwal, Brongersma, Eymery, Feiner,
  Forchel, Scheffler, Riess, Ohlsson, G\"{o}sele, and
  Samuelson]{ThelanderMatsTod06}
Thelander,~C.; Agarwal,~P.; Brongersma,~S.; Eymery,~J.; Feiner,~L.~F.;
  Forchel,~A.; Scheffler,~M.; Riess,~W.; Ohlsson,~B.~J.; G\"{o}sele,~U.;
  Samuelson,~L. \emph{Materials today} \textbf{2006}, \emph{9}, 28 -- 35\relax
\mciteBstWouldAddEndPuncttrue
\mciteSetBstMidEndSepPunct{\mcitedefaultmidpunct}
{\mcitedefaultendpunct}{\mcitedefaultseppunct}\relax
\EndOfBibitem
\bibitem[Armand(1986)]{ArmandAnnuRevMatSci86}
Armand,~M.~B. \emph{Annual Review of Materials Science} \textbf{1986},
  \emph{16}, 245--261\relax
\mciteBstWouldAddEndPuncttrue
\mciteSetBstMidEndSepPunct{\mcitedefaultmidpunct}
{\mcitedefaultendpunct}{\mcitedefaultseppunct}\relax
\EndOfBibitem
\bibitem[Takeya et~al.(2006)Takeya, Yamada, Hara, Shigeto, Tsukagoshi, Ikehata,
  and Aoyagi]{TakeyaAPL06}
Takeya,~J.; Yamada,~K.; Hara,~K.; Shigeto,~K.; Tsukagoshi,~K.; Ikehata,~S.;
  Aoyagi,~Y. \emph{Applied Physics Letters} \textbf{2006}, \emph{88},
  112102\relax
\mciteBstWouldAddEndPuncttrue
\mciteSetBstMidEndSepPunct{\mcitedefaultmidpunct}
{\mcitedefaultendpunct}{\mcitedefaultseppunct}\relax
\EndOfBibitem
\bibitem[Panzer et~al.(2005)Panzer, Newman, and Frisbie]{PanzerAPL05}
Panzer,~M.~J.; Newman,~C.~R.; Frisbie,~C.~D. \emph{Applied Physics Letters}
  \textbf{2005}, \emph{86}, 103503\relax
\mciteBstWouldAddEndPuncttrue
\mciteSetBstMidEndSepPunct{\mcitedefaultmidpunct}
{\mcitedefaultendpunct}{\mcitedefaultseppunct}\relax
\EndOfBibitem
\bibitem[Panzer and Frisbie(2006)Panzer, and Frisbie]{PanzerAFM06}
Panzer,~M.~J.; Frisbie,~C.~D. \emph{Advanced Functional Materials}
  \textbf{2006}, \emph{16}, 1051--1056\relax
\mciteBstWouldAddEndPuncttrue
\mciteSetBstMidEndSepPunct{\mcitedefaultmidpunct}
{\mcitedefaultendpunct}{\mcitedefaultseppunct}\relax
\EndOfBibitem
\bibitem[Rosenblatt et~al.(2002)Rosenblatt, Yaish, Park, Gore, Sazonova, and
  {McEuen}]{RosenblattNL02}
Rosenblatt,~S.; Yaish,~Y.; Park,~J.; Gore,~J.; Sazonova,~V.; {McEuen},~P.~L.
  \emph{Nano Letters} \textbf{2002}, \emph{2}, 869--872\relax
\mciteBstWouldAddEndPuncttrue
\mciteSetBstMidEndSepPunct{\mcitedefaultmidpunct}
{\mcitedefaultendpunct}{\mcitedefaultseppunct}\relax
\EndOfBibitem
\bibitem[Siddons et~al.(2004)Siddons, Merchin, Back, Jeong, and
  Shim]{SiddonsNL04}
Siddons,~G.~P.; Merchin,~D.; Back,~J.~H.; Jeong,~J.~K.; Shim,~M. \emph{Nano
  Letters} \textbf{2004}, \emph{4}, 927--931\relax
\mciteBstWouldAddEndPuncttrue
\mciteSetBstMidEndSepPunct{\mcitedefaultmidpunct}
{\mcitedefaultendpunct}{\mcitedefaultseppunct}\relax
\EndOfBibitem
\bibitem[Ozel et~al.(2005)Ozel, Gaur, Rogers, and Shim]{OzelNL05}
Ozel,~T.; Gaur,~A.; Rogers,~J.~A.; Shim,~M. \emph{Nano Letters} \textbf{2005},
  \emph{5}, 905--911\relax
\mciteBstWouldAddEndPuncttrue
\mciteSetBstMidEndSepPunct{\mcitedefaultmidpunct}
{\mcitedefaultendpunct}{\mcitedefaultseppunct}\relax
\EndOfBibitem
\bibitem[Liang and Gao(2012)Liang, and Gao]{LiangNL12}
Liang,~D.; Gao,~X. P.~A. \emph{Nano Letters} \textbf{2012}, \emph{12},
  3263--3267\relax
\mciteBstWouldAddEndPuncttrue
\mciteSetBstMidEndSepPunct{\mcitedefaultmidpunct}
{\mcitedefaultendpunct}{\mcitedefaultseppunct}\relax
\EndOfBibitem
\bibitem[Duan et~al.(2001)Duan, Huang, Cui, Wang, and Lieber]{DuanNat01}
Duan,~X.; Huang,~Y.; Cui,~Y.; Wang,~J.; Lieber,~C.~M. \emph{Nature}
  \textbf{2001}, \emph{409}, 66--69\relax
\mciteBstWouldAddEndPuncttrue
\mciteSetBstMidEndSepPunct{\mcitedefaultmidpunct}
{\mcitedefaultendpunct}{\mcitedefaultseppunct}\relax
\EndOfBibitem
\bibitem[Fasth et~al.(2005)Fasth, Fuhrer, Bj\"{o}rk, and Samuelson]{FasthNL05}
Fasth,~C.; Fuhrer,~A.; Bj\"{o}rk,~M.~T.; Samuelson,~L. \emph{Nano Letters}
  \textbf{2005}, \emph{5}, 1487--1490\relax
\mciteBstWouldAddEndPuncttrue
\mciteSetBstMidEndSepPunct{\mcitedefaultmidpunct}
{\mcitedefaultendpunct}{\mcitedefaultseppunct}\relax
\EndOfBibitem
\bibitem[Pfund et~al.(2006)Pfund, Shorubalko, Leturcq, and Ensslin]{PfundAPL06}
Pfund,~A.; Shorubalko,~I.; Leturcq,~R.; Ensslin,~K. \emph{Applied Physics
  Letters} \textbf{2006}, \emph{89}, 252106\relax
\mciteBstWouldAddEndPuncttrue
\mciteSetBstMidEndSepPunct{\mcitedefaultmidpunct}
{\mcitedefaultendpunct}{\mcitedefaultseppunct}\relax
\EndOfBibitem
\bibitem[Storm et~al.(2012)Storm, Nylund, Samuelson, and Micolich]{StormNL12}
Storm,~K.; Nylund,~G.; Samuelson,~L.; Micolich,~A.~P. \emph{Nano Letters}
  \textbf{2012}, \emph{12}, 1--6\relax
\mciteBstWouldAddEndPuncttrue
\mciteSetBstMidEndSepPunct{\mcitedefaultmidpunct}
{\mcitedefaultendpunct}{\mcitedefaultseppunct}\relax
\EndOfBibitem
\bibitem[Dhara et~al.(2011)Dhara, Sengupta, Solanki, Maurya, Pavan~R., Gokhale,
  Bhattacharya, and Deshmukh]{DharaAPL11}
Dhara,~S.; Sengupta,~S.; Solanki,~H.~S.; Maurya,~A.; Pavan~R.,~A.;
  Gokhale,~M.~R.; Bhattacharya,~A.; Deshmukh,~M.~M. \emph{Applied Physics
  Letters} \textbf{2011}, \emph{99}, 173101--173103\relax
\mciteBstWouldAddEndPuncttrue
\mciteSetBstMidEndSepPunct{\mcitedefaultmidpunct}
{\mcitedefaultendpunct}{\mcitedefaultseppunct}\relax
\EndOfBibitem
\bibitem[Khanal and Wu(2007)Khanal, and Wu]{KhanalNL07}
Khanal,~D.~R.; Wu,~J. \emph{Nano Letters} \textbf{2007}, \emph{7},
  2778--2783\relax
\mciteBstWouldAddEndPuncttrue
\mciteSetBstMidEndSepPunct{\mcitedefaultmidpunct}
{\mcitedefaultendpunct}{\mcitedefaultseppunct}\relax
\EndOfBibitem
\bibitem[Thelander et~al.(2008)Thelander, Rehnstedt, Fr\"{o}berg, Lind,
  Martensson, Caroff, Lowgren, Ohlsson, Samuelson, and
  Wernersson]{ThelanderTED08}
Thelander,~C.; Rehnstedt,~C.; Fr\"{o}berg,~L.~E.; Lind,~E.; Martensson,~T.;
  Caroff,~P.; Lowgren,~T.; Ohlsson,~B.~J.; Samuelson,~L.; Wernersson,~L.-E.
  \emph{{IEEE} Transactions on Electron Devices} \textbf{2008}, \emph{55},
  3030--3036\relax
\mciteBstWouldAddEndPuncttrue
\mciteSetBstMidEndSepPunct{\mcitedefaultmidpunct}
{\mcitedefaultendpunct}{\mcitedefaultseppunct}\relax
\EndOfBibitem
\bibitem[Ford et~al.(2012)Ford, Kumar, Kapadia, Guo, and Javey]{FordNL12}
Ford,~A.~C.; Kumar,~S.~B.; Kapadia,~R.; Guo,~J.; Javey,~A. \emph{Nano Letters}
  \textbf{2012}, \emph{12}, 1340--1343\relax
\mciteBstWouldAddEndPuncttrue
\mciteSetBstMidEndSepPunct{\mcitedefaultmidpunct}
{\mcitedefaultendpunct}{\mcitedefaultseppunct}\relax
\EndOfBibitem
\bibitem[Tian et~al.(2012)Tian, Sakr, Kinder, Liang, {MacDonald}, Qiu, Gao, and
  Gao]{TianNL12}
Tian,~Y.; Sakr,~M.~R.; Kinder,~J.~M.; Liang,~D.; {MacDonald},~M.~J.; Qiu,~R.
  L.~J.; Gao,~H.-J.; Gao,~X. P.~A. \emph{Nano Letters} \textbf{2012},
  \emph{12}, 6492--6497\relax
\mciteBstWouldAddEndPuncttrue
\mciteSetBstMidEndSepPunct{\mcitedefaultmidpunct}
{\mcitedefaultendpunct}{\mcitedefaultseppunct}\relax
\EndOfBibitem
\bibitem[van Weperen et~al.(2013)van Weperen, Plissard, Bakkers, Frolov, and
  Kouwenhoven]{vanWeperenNL13}
van Weperen,~I.; Plissard,~S.~R.; Bakkers,~E. P. A.~M.; Frolov,~S.~M.;
  Kouwenhoven,~L.~P. \emph{Nano Letters} \textbf{2013}, \emph{13},
  387--391\relax
\mciteBstWouldAddEndPuncttrue
\mciteSetBstMidEndSepPunct{\mcitedefaultmidpunct}
{\mcitedefaultendpunct}{\mcitedefaultseppunct}\relax
\EndOfBibitem
\bibitem[de~Gans et~al.(2004)de~Gans, Duineveld, and Schubert]{deGansAdvMat04}
de~Gans,~B.-J.; Duineveld,~P.~C.; Schubert,~U.~S. \emph{Advanced Materials}
  \textbf{2004}, \emph{16}, 203--213\relax
\mciteBstWouldAddEndPuncttrue
\mciteSetBstMidEndSepPunct{\mcitedefaultmidpunct}
{\mcitedefaultendpunct}{\mcitedefaultseppunct}\relax
\EndOfBibitem
\bibitem[Cho et~al.(2008)Cho, Lee, Xia, Kim, He, Renn, Lodge, and
  Frisbie]{ChoNatMat08}
Cho,~J.~H.; Lee,~J.; Xia,~Y.; Kim,~B.; He,~Y.; Renn,~M.~J.; Lodge,~T.~P.;
  Frisbie,~C.~D. \emph{Nature Materials} \textbf{2008}, \emph{7},
  900--906\relax
\mciteBstWouldAddEndPuncttrue
\mciteSetBstMidEndSepPunct{\mcitedefaultmidpunct}
{\mcitedefaultendpunct}{\mcitedefaultseppunct}\relax
\EndOfBibitem
\bibitem[Choi et~al.(2010)Choi, Lee, Kar, Das, Jeon, Moon, Lee, Jeong, and
  Myoung]{ChoiJMatChem10}
Choi,~J.-H.; Lee,~S.~W.; Kar,~J.~P.; Das,~S.~N.; Jeon,~J.; Moon,~K.-J.;
  Lee,~T.~I.; Jeong,~U.; Myoung,~J.-M. \emph{Journal of Materials Chemistry}
  \textbf{2010}, \emph{20}, 7393--7397\relax
\mciteBstWouldAddEndPuncttrue
\mciteSetBstMidEndSepPunct{\mcitedefaultmidpunct}
{\mcitedefaultendpunct}{\mcitedefaultseppunct}\relax
\EndOfBibitem
\bibitem[Krsko et~al.(2003)Krsko, Sukhishvili, Mansfield, Clancy, and
  Libera]{KrskoLangmuir03}
Krsko,~P.; Sukhishvili,~S.; Mansfield,~M.; Clancy,~R.; Libera,~M.
  \emph{Langmuir} \textbf{2003}, \emph{19}, 5618--5625\relax
\mciteBstWouldAddEndPuncttrue
\mciteSetBstMidEndSepPunct{\mcitedefaultmidpunct}
{\mcitedefaultendpunct}{\mcitedefaultseppunct}\relax
\EndOfBibitem
\bibitem[Hong et~al.(2004)Hong, Krsko, and Libera]{HongLangmuir04}
Hong,~Y.; Krsko,~P.; Libera,~M. \emph{Langmuir} \textbf{2004}, \emph{20},
  11123--11126\relax
\mciteBstWouldAddEndPuncttrue
\mciteSetBstMidEndSepPunct{\mcitedefaultmidpunct}
{\mcitedefaultendpunct}{\mcitedefaultseppunct}\relax
\EndOfBibitem
\bibitem[Uchiyama et~al.(2009)Uchiyama, Kusagawa, Hanai, Imanishi, Hirano, and
  Takeda]{UchiyamaSSI09}
Uchiyama,~R.; Kusagawa,~K.; Hanai,~K.; Imanishi,~N.; Hirano,~A.; Takeda,~Y.
  \emph{Solid State Ionics} \textbf{2009}, \emph{180}, 205--211\relax
\mciteBstWouldAddEndPuncttrue
\mciteSetBstMidEndSepPunct{\mcitedefaultmidpunct}
{\mcitedefaultendpunct}{\mcitedefaultseppunct}\relax
\EndOfBibitem
\bibitem[Ueno et~al.(2011)Ueno, Imanishi, Hanai, Kobayashi, Hirano, Yamamoto,
  and Takeda]{UenoJPS11}
Ueno,~M.; Imanishi,~N.; Hanai,~K.; Kobayashi,~T.; Hirano,~A.; Yamamoto,~O.;
  Takeda,~Y. \emph{Journal of Power Sources} \textbf{2011}, \emph{196},
  4756--4761\relax
\mciteBstWouldAddEndPuncttrue
\mciteSetBstMidEndSepPunct{\mcitedefaultmidpunct}
{\mcitedefaultendpunct}{\mcitedefaultseppunct}\relax
\EndOfBibitem
\bibitem[Murata et~al.(1981)Murata, Kyser, and Ting]{MurataJAP81}
Murata,~K.; Kyser,~D.~F.; Ting,~C.~H. \emph{Journal of Applied Physics}
  \textbf{1981}, \emph{52}, 4396--4405\relax
\mciteBstWouldAddEndPuncttrue
\mciteSetBstMidEndSepPunct{\mcitedefaultmidpunct}
{\mcitedefaultendpunct}{\mcitedefaultseppunct}\relax
\EndOfBibitem
\bibitem[Shimizu and Ze-Jun(1992)Shimizu, and Ze-Jun]{ShimizuRPP92}
Shimizu,~R.; Ze-Jun,~D. \emph{Reports on Progress in Physics} \textbf{1992},
  \emph{55}, 487--531\relax
\mciteBstWouldAddEndPuncttrue
\mciteSetBstMidEndSepPunct{\mcitedefaultmidpunct}
{\mcitedefaultendpunct}{\mcitedefaultseppunct}\relax
\EndOfBibitem
\bibitem[Aizaki(1979)]{AizakiJVST79}
Aizaki,~N. \emph{Journal of Vacuum Science and Technology} \textbf{1979},
  \emph{16}, 1726--1733\relax
\mciteBstWouldAddEndPuncttrue
\mciteSetBstMidEndSepPunct{\mcitedefaultmidpunct}
{\mcitedefaultendpunct}{\mcitedefaultseppunct}\relax
\EndOfBibitem
\bibitem[Lind et~al.(2006)Lind, Persson, Samuelson, and Wernersson]{LindNL06}
Lind,~E.; Persson,~A.~I.; Samuelson,~L.; Wernersson,~L.-E. \emph{Nano Letters}
  \textbf{2006}, \emph{6}, 1842--1846\relax
\mciteBstWouldAddEndPuncttrue
\mciteSetBstMidEndSepPunct{\mcitedefaultmidpunct}
{\mcitedefaultendpunct}{\mcitedefaultseppunct}\relax
\EndOfBibitem
\bibitem[Panzer and Frisbie(2006)Panzer, and Frisbie]{PanzerAPL06}
Panzer,~M.~J.; Frisbie,~C.~D. \emph{Applied Physics Letters} \textbf{2006},
  \emph{88}, 203504--203506\relax
\mciteBstWouldAddEndPuncttrue
\mciteSetBstMidEndSepPunct{\mcitedefaultmidpunct}
{\mcitedefaultendpunct}{\mcitedefaultseppunct}\relax
\EndOfBibitem
\bibitem[Bruce and Vincent(1993)Bruce, and Vincent]{BruceJChemSocFarTran93}
Bruce,~P.~G.; Vincent,~C.~A. \emph{J. Chem. Soc., Faraday Trans.}
  \textbf{1993}, \emph{89}, 3187 -- 3203\relax
\mciteBstWouldAddEndPuncttrue
\mciteSetBstMidEndSepPunct{\mcitedefaultmidpunct}
{\mcitedefaultendpunct}{\mcitedefaultseppunct}\relax
\EndOfBibitem
\bibitem[{Fullerton-Shirey} and Maranas(2009){Fullerton-Shirey}, and
  Maranas]{Fullerton-ShireyMacroMol09}
{Fullerton-Shirey},~S.~K.; Maranas,~J.~K. \emph{Macromolecules} \textbf{2009},
  \emph{42}, 2142--2156\relax
\mciteBstWouldAddEndPuncttrue
\mciteSetBstMidEndSepPunct{\mcitedefaultmidpunct}
{\mcitedefaultendpunct}{\mcitedefaultseppunct}\relax
\EndOfBibitem
\bibitem[Rehnstedt et~al.(2008)Rehnstedt, Martensson, Thelander, Samuelson, and
  Wernersson]{RehnstedtTED08}
Rehnstedt,~C.; Martensson,~T.; Thelander,~C.; Samuelson,~L.; Wernersson,~L.-E.
  \emph{{IEEE} Transactions on Electron Devices} \textbf{2008}, \emph{55},
  3037--3041\relax
\mciteBstWouldAddEndPuncttrue
\mciteSetBstMidEndSepPunct{\mcitedefaultmidpunct}
{\mcitedefaultendpunct}{\mcitedefaultseppunct}\relax
\EndOfBibitem
\bibitem[Fr\"{o}berg et~al.(2008)Fr\"{o}berg, Rehnstedt, Thelander, Lind,
  Wernersson, and Samuelson]{FrobergEDL08}
Fr\"{o}berg,~L.~E.; Rehnstedt,~C.; Thelander,~C.; Lind,~E.; Wernersson,~L.;
  Samuelson,~L. \emph{{IEEE} Electron Device Letters} \textbf{2008}, \emph{29},
  981--983\relax
\mciteBstWouldAddEndPuncttrue
\mciteSetBstMidEndSepPunct{\mcitedefaultmidpunct}
{\mcitedefaultendpunct}{\mcitedefaultseppunct}\relax
\EndOfBibitem
\bibitem[Rehnstedt et~al.(2008)Rehnstedt, Thelander, Fr\"{o}berg, Ohlsson,
  Samuelson, and Wernersson]{RehnstedtEL08}
Rehnstedt,~C.; Thelander,~C.; Fr\"{o}berg,~L.~E.; Ohlsson,~B.~J.;
  Samuelson,~L.; Wernersson,~L.-E. \emph{Electronics Letters} \textbf{2008},
  \emph{44}, 236\relax
\mciteBstWouldAddEndPuncttrue
\mciteSetBstMidEndSepPunct{\mcitedefaultmidpunct}
{\mcitedefaultendpunct}{\mcitedefaultseppunct}\relax
\EndOfBibitem
\bibitem[Tanaka et~al.(2010)Tanaka, Tomioka, Hara, Motohisa, Sano, and
  Fukui]{TanakaAPEX10}
Tanaka,~T.; Tomioka,~K.; Hara,~S.; Motohisa,~J.; Sano,~E.; Fukui,~T.
  \emph{Applied Physics Express} \textbf{2010}, \emph{3}, 025003\relax
\mciteBstWouldAddEndPuncttrue
\mciteSetBstMidEndSepPunct{\mcitedefaultmidpunct}
{\mcitedefaultendpunct}{\mcitedefaultseppunct}\relax
\EndOfBibitem
\bibitem[Roddaro et~al.(2011)Roddaro, Pescaglini, Ercolani, Sorba, and
  Beltram]{RoddaroNL11}
Roddaro,~S.; Pescaglini,~A.; Ercolani,~D.; Sorba,~L.; Beltram,~F. \emph{Nano
  Letters} \textbf{2011}, \emph{11}, 1695--1699\relax
\mciteBstWouldAddEndPuncttrue
\mciteSetBstMidEndSepPunct{\mcitedefaultmidpunct}
{\mcitedefaultendpunct}{\mcitedefaultseppunct}\relax
\EndOfBibitem
\bibitem[Croce et~al.(1998)Croce, Appetecchi, Persi, and Scrosati]{CroceNat98}
Croce,~F.; Appetecchi,~G.~B.; Persi,~L.; Scrosati,~B. \emph{Nature}
  \textbf{1998}, \emph{394}, 456\relax
\mciteBstWouldAddEndPuncttrue
\mciteSetBstMidEndSepPunct{\mcitedefaultmidpunct}
{\mcitedefaultendpunct}{\mcitedefaultseppunct}\relax
\EndOfBibitem
\bibitem[Qian et~al.(2001)Qian, Gu, Cheng, Yang, Wang, and Dong]{QianJSSE01}
Qian,~X.; Gu,~N.; Cheng,~Z.; Yang,~X.; Wang,~E.; Dong,~S. \emph{Journal of
  Solid State Electrochemistry} \textbf{2001}, \emph{6}, 8--15\relax
\mciteBstWouldAddEndPuncttrue
\mciteSetBstMidEndSepPunct{\mcitedefaultmidpunct}
{\mcitedefaultendpunct}{\mcitedefaultseppunct}\relax
\EndOfBibitem
\bibitem[Wang et~al.(2005)Wang, Yang, Wang, Liu, and
  Dou]{WangJNanosciNanotech05}
Wang,~G.~X.; Yang,~L.; Wang,~J.~Z.; Liu,~H.~K.; Dou,~S.~X. \emph{Journal of
  Nanoscience and Nanotechnology} \textbf{2005}, \emph{5}, 1135--1140\relax
\mciteBstWouldAddEndPuncttrue
\mciteSetBstMidEndSepPunct{\mcitedefaultmidpunct}
{\mcitedefaultendpunct}{\mcitedefaultseppunct}\relax
\EndOfBibitem
\bibitem[Klingshirn et~al.(2004)Klingshirn, Spear, Subramanian, Holbrey,
  Huddleston, and Rogers]{KlingshirnChemMat04}
Klingshirn,~M.~A.; Spear,~S.~K.; Subramanian,~R.; Holbrey,~J.~D.;
  Huddleston,~J.~G.; Rogers,~R.~D. \emph{Chemistry of Materials} \textbf{2004},
  \emph{16}, 3091--3097\relax
\mciteBstWouldAddEndPuncttrue
\mciteSetBstMidEndSepPunct{\mcitedefaultmidpunct}
{\mcitedefaultendpunct}{\mcitedefaultseppunct}\relax
\EndOfBibitem
\bibitem[Susan et~al.(2005)Susan, Kaneko, Noda, and Watanabe]{SusanJACS05}
Susan,~M. A. B.~H.; Kaneko,~T.; Noda,~A.; Watanabe,~M. \emph{Journal of the
  American Chemical Society} \textbf{2005}, \emph{127}, 4976--4983\relax
\mciteBstWouldAddEndPuncttrue
\mciteSetBstMidEndSepPunct{\mcitedefaultmidpunct}
{\mcitedefaultendpunct}{\mcitedefaultseppunct}\relax
\EndOfBibitem
\bibitem[Lee et~al.(2007)Lee, Panzer, He, Lodge, and Frisbie]{LeeJACS07}
Lee,~J.; Panzer,~M.~J.; He,~Y.; Lodge,~T.~P.; Frisbie,~C.~D. \emph{Journal of
  the American Chemical Society} \textbf{2007}, \emph{129}, 4532--4533\relax
\mciteBstWouldAddEndPuncttrue
\mciteSetBstMidEndSepPunct{\mcitedefaultmidpunct}
{\mcitedefaultendpunct}{\mcitedefaultseppunct}\relax
\EndOfBibitem
\bibitem[Zailer et~al.(1996)Zailer, Frost, {Chabasseur-Molyneux}, Ford, and
  Pepper]{ZailerSST96}
Zailer,~I.; Frost,~J. E.~F.; {Chabasseur-Molyneux},~V.; Ford,~C. J.~B.;
  Pepper,~M. \emph{Semiconductor Science and Technology} \textbf{1996},
  \emph{11}, 1235--1238\relax
\mciteBstWouldAddEndPuncttrue
\mciteSetBstMidEndSepPunct{\mcitedefaultmidpunct}
{\mcitedefaultendpunct}{\mcitedefaultseppunct}\relax
\EndOfBibitem
\bibitem[Suyatin et~al.(2007)Suyatin, Thelander, Bj\"{o}rk, Maximov, and
  Samuelson]{SuyatinNanotech07}
Suyatin,~D.~B.; Thelander,~C.; Bj\"{o}rk,~M.~T.; Maximov,~I.; Samuelson,~L.
  \emph{Nanotechnology} \textbf{2007}, \emph{18}, 105307\relax
\mciteBstWouldAddEndPuncttrue
\mciteSetBstMidEndSepPunct{\mcitedefaultmidpunct}
{\mcitedefaultendpunct}{\mcitedefaultseppunct}\relax
\EndOfBibitem
\end{mcitethebibliography}

\end{document}